\documentclass[A4paper,pre,twocolumn,showpacs,aps,floats,floatfix,superscriptaddress]{revtex4}

\usepackage{graphicx}
\usepackage{graphics}
\usepackage{subfigure}
\usepackage{epsfig}
\usepackage{amsmath,amssymb,bm}

\usepackage[colorlinks,linkcolor=blue,citecolor=blue]{hyperref}
\usepackage{dcolumn}
\usepackage{rotating}
\usepackage{multirow}
\usepackage{epsfig}
\usepackage{color}
\usepackage{float}
\usepackage{setspace}
\usepackage{natbib}

\begin{document}

\title{Voting Behavior, Coalitions and Government Strength through a Complex Network Analysis}

\author{Carlo Dal Maso} \affiliation{IMT, Institute for Advanced Studies, Lucca, Italy}  
\author{Gabriele Pompa} \affiliation{IMT, Institute for Advanced Studies, Lucca, Italy} 
\author{Michelangelo Puliga} \affiliation{IMT, Institute for Advanced Studies, Lucca, Italy} \affiliation{Linkalab, Complex
  Systems Computational Laboratory, Cagliari, Italy}
\author{Gianni Riotta} \affiliation{IMT, Institute for Advanced Studies, Lucca, Italy} \affiliation{Department of French and Italian, Pirelli Chair, Princenton Univerity, New Jersey, USA}
\author{Alessandro Chessa} \affiliation{IMT, Institute for Advanced Studies, Lucca, Italy} \affiliation{Linkalab, Complex
  Systems Computational Laboratory, Cagliari, Italy}

\date{\today}

%\widetext
\begin{abstract}

  We analyze the network of relations between parliament members according to
their voting behavior.  In particular, we examine the emergent community
structure with respect to political coalitions and government alliances. We rely
on tools developed in the Complex Network literature to explore the core of
these communities and use their topological features to develop new metrics for party polarization, internal coalition cohesiveness and government strength. As a case study, we focus on the  Chamber of Deputies of the Italian Parliament, for which we are able to characterize the heterogeneity of the ruling coalition as well as parties� specific contributions to the stability of the government over time. We find sharp contrast in the political debate which surprisingly does not imply a relevant structure based on establised parties. We take a closer look to changes in the community structure after parties split up and their effect on the position of single deputies within communities. Finally, we introduce a way to track the stability of the government coalition over time that is able to discern the contribution of each member along with the impact of its possible defection. While our case study relies on the Italian parliament, whose relevance has come into the international spotlight in the present economic downturn, the methods developed here are entirely general and can therefore be applied to a multitude of other scenarios.

\end{abstract}

%\pacs{89.75.-k, -87.23.Ge, 05.70Ln}

\maketitle

%\section*{Author Summary}

\section{Introduction}
A great deal of recent research has been devoted to explaining political
polarization in parliaments \cite{c1, c2}. This literature has been dominated by
models where party polarization is either explained by the polarization of the
electorate or through the party and ideology of deputies. A new stem of
literature has recently adopted tools of Complex Network Science
\cite{Caldarelli2007Scale-FreeNetworksComplex,newman2003structure} to
investigate this issue, with a network representation being given to committees and subcommittees who share the
same members in the US Congress \cite{c3}, to members of the Congress who
co-sponsor bills \cite{c4,c5} or those who place the same roll-call votes
\cite{c6}. We follow the latter approach so that deputies are represented as nodes within a network where the number
of shared roll-call votes determines the strength of their links. Similarly to \cite{c7, c8} we make use of the
network science concept of �modularity� in order to reconstruct the community
structure of the parliament \cite{c11}. We introduce a novel method to
characterize the position of each deputy in the community of reference, based on
its contribution to the modularity score, proposing a more intuitive
interpretation compared to that based on the spectral decomposition developed in \cite{c8} and in \cite{c3}. The method
presented here can be easily generalized on a wider European scale, and
replicated across a longer time span or in industry-specific policies. The rest
of the paper is organized as follows. In the ``Methods'' section we present
the methodology used to investigate parliamentary polarization, party cohesion, community structure and their time
evolution, in the ``Results" section the main findings related to the specific case of the
Italian Parliament are presented, while in the ``Discussion'' section we draw our conclusions and sketch the lines of future research.

\section{Methods}
As the first step in our methodology we construct a graph where each node
represents one of the $n$ deputies and edges are drawn every time two deputies display the same voting behavior\footnote{i.e. both vote in favor, against or abstain from vote. No edges are drawn for absent deputies.}.
We then normalize edges by the total number of votes in the reference period in
order to obtain a weighted graph where weights are $0 \leq w_{ij} \leq 1$. Full weight is given to two deputies $i,j$ if they participated in all sessions and voted exactly the same way in all of sessions. When a deputy quits the parliament, because of incompatibility, resignation etc., and his or her seat is taken by a new person, we consider the two deputies as being just one node\footnote{We check whether this transition leads to some votes in which none of the two deputies had their chairs without finding any discontinuity.}.\\
Initially, we look at the topological structure of parties in order to study their cohesion over time. Completely ignoring any \emph{a priori} knowledge of party affiliation, we look at the communities arising directly from voting behavior to see whether they match or not.

\subsection*{Analysis of party cohesion}
Consider each party as a subgraph $C$ of the graph $G$, with $n_{C}$ being the number of deputies in the party. An intuitive way of measuring party cohesion (i.e. the tendency of the party to vote as a single entity) is to evaluate the intra-cluster density $d_{int}(C)$ defined as the ratio between the total internal strength of the sub-graph $C$ and the number of all possible edges inside that cluster \cite{c10}
$$
d_{int}(C)=\frac{\sum_{ij \in C} w_{ij}}{n_{c} \left( n_{c} - 1\right)/2} \mbox{ }. 
$$
Similarly, we can define the inter-cluster density $d_{ext}(C)$ as the ratio between the observed strength of edges running from the nodes of $C$ to the rest of the network and the maximum number of edges connecting internal with external nodes:
$$
d_{ext}(C)=\frac{\sum_{i \in C, j \notin C} w_{ij}}{n_{c} \left(n -  n_{c}\right)} \mbox{ }. 
$$
A party stands out as a specific political group if $d_{int}(C)$ is appreciably larger than the average link density $d(G)=\sum_{i,j}w_{ij}/\frac{n(n-1)}{2}$ of the entire network $G$ and similarly we expect $d_{ext}(C)$ to be appreciably smaller.\\ 
Searching for the best tradeoff between a large intra-cluster density and a small inter-cluster one is indeed an implicit or explicit goal for most algorithms used in community detection \cite{c10, c11}.\\

\subsection*{Community and core detection}
Modularity optimization is a well-established method for detecting communities
\cite{c11}. The idea behind modularity is that a random graph should not have a
cluster structure so that communities are revealed maximizing the difference
between the density of edges in a sub-graph and that expected if edges were
connected at random. Hence the modularity function of a weighted graph 
\cite{newman2004analysis}, where in our case nodes are deputies and edges
represent the percentage of votes that two of them have in common, is given by:
$$
Q=\frac{1}{2W}\sum_{i,j}\left(A_{ij}-\frac{s_{i}s_{j}}{2W}\right)\delta\left(C_{i}, C_{j} \right)
$$ 
where $A_{ij}$ gives the fraction of similar votes deputies $i$ and $j$ share in
common ($A_{ij}=w_{ij}$), $W$ is the total weight in the network, $\delta\left(C_{i}, C_{j} \right)$ is a delta function that yields one if deputies $i$ and $j$ are in the same community ($C_{i}=C_{j}$) and $0$ otherwise, and $s_{i}, s_{j}$ represent the strength of node $i$ and $j$ respectively.\\
In the general case of modern democracies the typical result of the modularity
optimization should be the splitting of the graph into two communities that
reproduces the government coalition and the opposition.\\
Moreover each node in its community usually doesn't have the same importance for
its stability. Indeed, the removal of a node in the community �core� should
affect the partition much more than the deletion of a boundary node.
In other terms, some deputies display such a high degree of internal connections
so that they can be identified as the bulk of the coalition. As we proceed
toward the boundary, deputies display increasing connections to the opposite community.\\
In order to investigate this structure, we exploit the properties of the
modularity function following a new procedure introduced in \cite{c12}. By
definition, if the modularity associated with a network has been optimized,
every perturbation of the partition leads to a negative variation of the
modularity $dQ < 0$.\\
We compute the effect on the modularity associated with the shift of a deputy
from one community to another and we plot the corresponding $dQ's$ distribution
in order to check the coreness of each deputy and his party.
In case of three or more communities $dQ$ was originally developed in \cite{c12} to report the minimum variation in modularity, i.e. modularity was compared against a setup in which each node was moved, one per time, to its \textit{closest} community.
Here we rather consider movements to the \textit{farthest} community in order to
avoid abrupt shifts in the distribution of $dQ$ due to the rise of small temporary (third) communities.
Finally the histogram of the $dQ's$ will highlight the different groups that
make up the coalition and will show different sub distributions along the
support interval of $dQ$.

\subsection*{Measures of Polarization, Cohesion  and Stability}
Dealing with roll-call vote's networks as a whole, standard approaches \cite{c6, c8, c9} have adopted the modularity score as a measure for party polarization. 
However, our methodology gives us the possibility to consider the overall voting
behavior on a much finer scale, considering the contribution of every single
deputy. In line with this, we have decided to measure the polarization as an
average decrease in the modularity score consequent to the substitution of two opposite deputies; the larger the decrease in the modularity score, the larger the current contraposition between the two coalitions becomes.
So we define the polarization as the
sum of the median of the monthly $dQ$ distributions of the two communities\footnote{The median has been preferred to the mean as a measure of location, because the distributions of both communities are strongly (negatively) skewed.}.

When we focus on features of only one community, we still need to account for
the community structure of the whole graph. Think for instance of two time
frames in which the members of the ruling coalition vote exactly the same while the opposition voted $1/2$ and $1/4$ of the time with the government. Then the government $dQ$ distribution would present more extreme values in the latter case, determining a shift towards more negative values of the mass of the entire distribution, despite the cohesion of the government \textit{per se} not changing at all.
Therefore any measure of cohesion should be robust to changes in the location of the distribution.
A suitable one is represented by the interquartile difference of the $dQ's$
distribution that we will employ as our standard definition for the
party/coalition cohesiveness.

In addition to polarization and cohesion, the stability of the government is
directly affected by the number of its loyal deputies; in order for laws to be
passed, half plus one of the total number of deputies are needed in the Chamber
of Deputies. So as a rough rule of thumb, we can consider a government that
keeps up to half plus one deputies on his side to still be \textit{safe}. 
This measure accounts for the stability of the government comunity in the shape
of a safety zone that divides the last critical deputy able to break down the
majority from the $dQ = 0$ postion before the oppositon community region.

\section{Results}
\begin{figure}[!ht]
\begin{center}
\hspace*{-0.6cm}\includegraphics[width=3.8in]{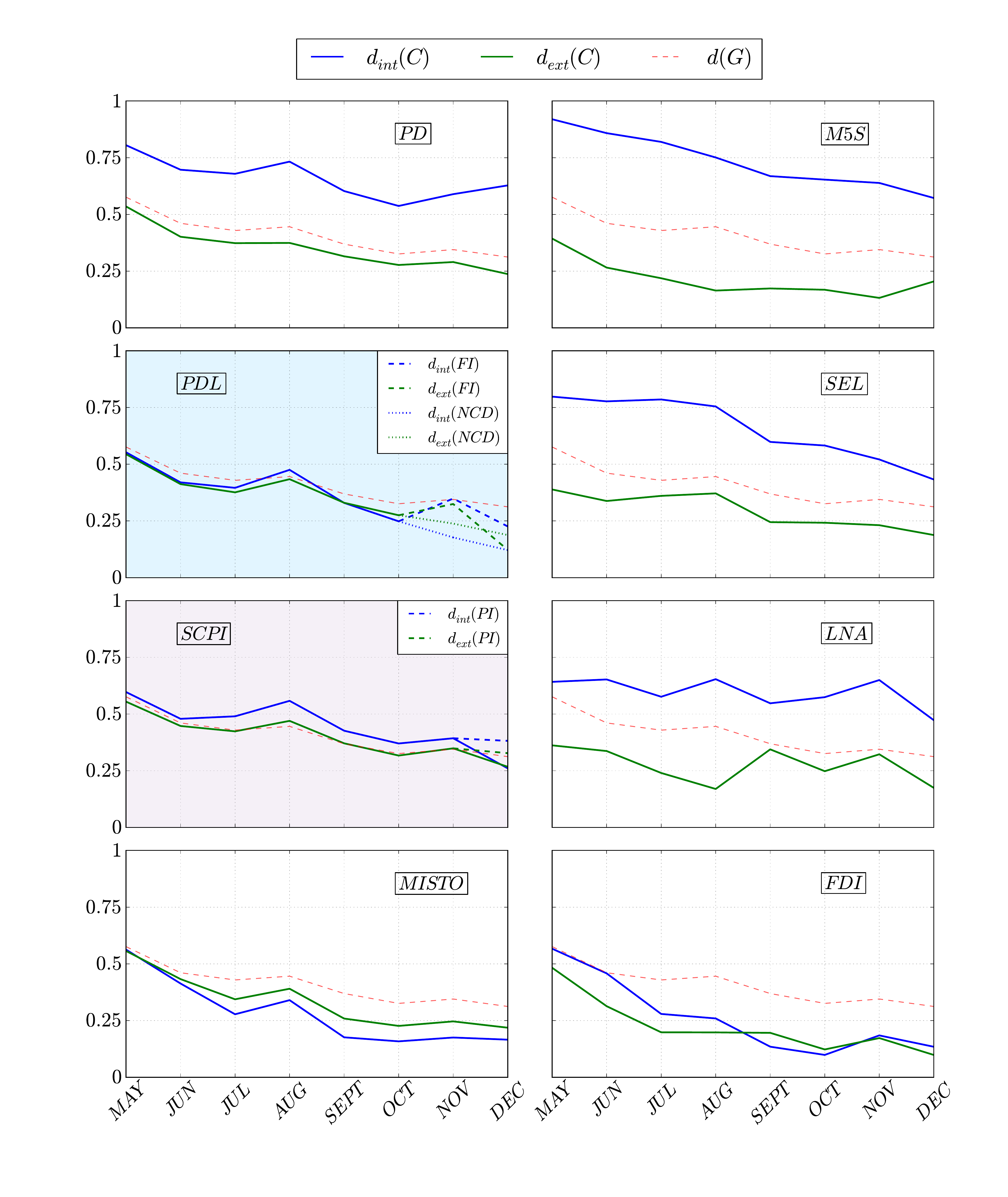}
\end{center}
\caption{
{\bf Members of a party show cohesion if the links connecting them are stronger than the ties with other deputies.} We capture the former by the intra-cluster density $d_{int}(C)$ and the latter by the inter-cluster density $d_{ext}(C)$. The party shows high cohesion when the two lie considerably higher and lower respectively copared to the average link density of the whole parliament $d(C)$.
}
\label{Figure_1}
\end{figure}

As a concrete case we analyze the network of deputies in the newly elected
Italian Chamber of Deputies (2013). We collect information on the 630 deputies
and their voting behavior from the government open data
SPARQL endpoint\footnote{http://dati.camera.it/sparql}. The reader may refer to
table \ref{tabella_carlo} for an outline of the main Italian parties mentioned in this paper.
The available data cover parliamentary votes from April 2013, when the new parliament was appointed, to the end of December 2013. Despite being quite a short period of time, the dataset covers 2820 parliamentary votes, which implies more than 1,5 million individual votes in our time span. Importantly, the Italian government has made semantic data following W3C standards available, which translates into fast and precise data manipulation through computer based queries. We refer to this source of data for the profiles of deputies and the classification of votes, while data on voting behavior of single deputies was taken directly from institutional web sites\footnote{http://documenti.camera.it/votazioni/votazionitutte /FormVotazioni.Asp?Legislatura=XVII}.\\
Figure \ref{Figure_1} represents the evolution of density measures over time for
each party in the Italian Chamber of Deputies. While the structures of the M5S,
PD, SEL and LNA parties are recognizable within the graph, the other groups
present inter- and intra-cluster densities that are very close to each other, or
at times even overlapping.

\begin{figure}[!ht]
\begin{center}
\hspace*{-0.7cm}\includegraphics[width=4in]{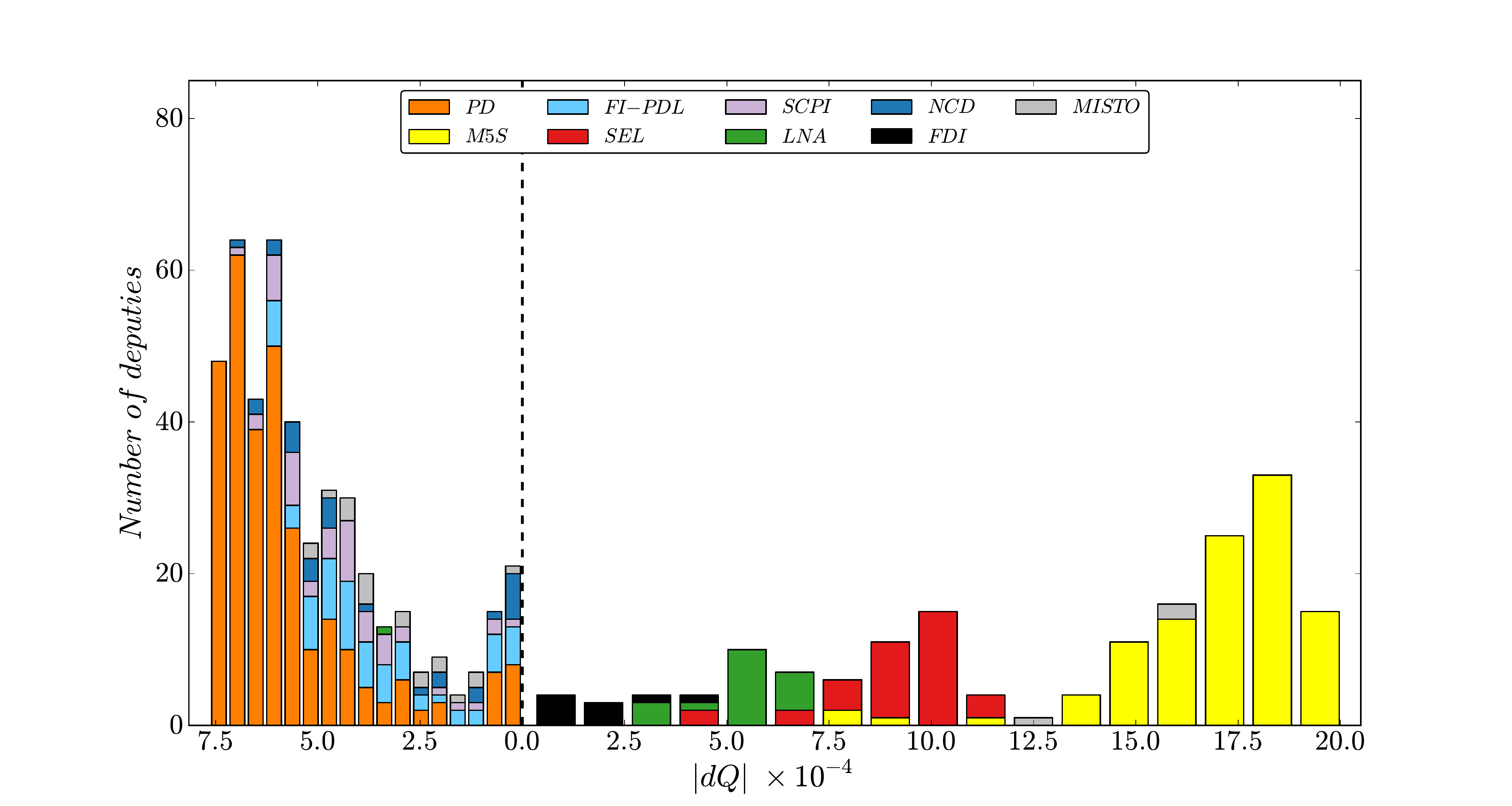}
\end{center}
\caption{
{\bf Community Structure  of the Italian Parliament.} The vertical dashed line separates the two main coalitions/communities (Government/Opposition). Each coalition comprises different parties corresponding to different colors.
The quantity 'dQ' is associated to the coreness of each deputy/party. The
distributions are obtained computing the coreness of each deputy and then
aggregating them in the form of a stacked istogram. The more the distance of the
bars from the vertical dashed line, the more the deputies/parties
are at the core of their coalition. Notice how the main parties tend to
segregate in clusterd distributions with different positions in the 'dQ'
axes.
}
\label{Figure_2}
\end{figure}

This means that at a certain point their votes
proportionally coincide to a greater degree with other groups than with their
own members. The plots marked with a colored background report the splitting of
two political groups, when the PDL breaks up into the NCD and the FI in November
2013 and the PI exits from the SCPI in December 2013. The inter-cluster density,
represented in green, is clearly higher for groups who support the government
(PD, PDL and SCPI). Theoretically these groups should vote in compliance with
the majority's prescriptions, thereby showing a similar voting behavior. Once we
take into account the average monthly levels of edge density $d(G)$ the
topological structure of parties becomes very similar to the rest of the graph.
As such, parties may not be the most appropriate representation of voting
structure, thus leading us to consider the behavioral identification of
political groups through the modularity function.
Once applied to the graph of deputies, the modularity optimization usually
splits the graph into two communities that almost exactly match the government
coalition and the opposition as shown in figure \ref{Figure_2} where the
vertical dashed line separates the two coalitions.

Afterword we compute the effect on the modularity associated with the removal
of a deputy from his community computing the corresponding $dQ's$ and the result
is also shown in figure \ref{Figure_2}.
The histogram shows the $dQ's$ distribution of the government's coalition on the
left side of the dashed line and that of the opposition on the right, with
alle the dQ associated to different parties in different colors.\\
Indeed, the core of the coalition appears to be made up by a relatively higher share of deputies from the center-left party PD while relatively more deputies from the center-right party PDL appear to be at the periphery as we move to the right. 
This provides an interesting insight on the rather different roles played by the two main Italian parties joined by a coalition pact, namely the PD and the PDL, with the latter ultimately quitting the government in mid November 2013. As for the opposition, note that the support of $dQ$ is far more dispersed with each group taking on a limited range of values in the  distribution. 
This is not surprising in that the opposition is not a coalition \textit{per se} but rather a set of groups that might vote with the ruling coalition depending on the subject at hand. In particular, deputies from the M5S make up the core of the opposition with a higher magnitude of $dQ$, which also holds true when compared to the core of the government coalition. 
This may be due to a relatively inflexible opposition to the government or in equal measure to the fact that it is the largest group in the opposition community. 
On the other hand, the SEL and the LNA are progressively closer to the border of the community, which may be reasonable if we consider that these groups used to be allies of two parties in the government coalition, namely the PD and the PDL respectively.

\subsection*{Time evolution of the community structure}
\begin{figure}[!ht]
\begin{center}
\hspace*{-0.5cm}\includegraphics[width=3.9in]{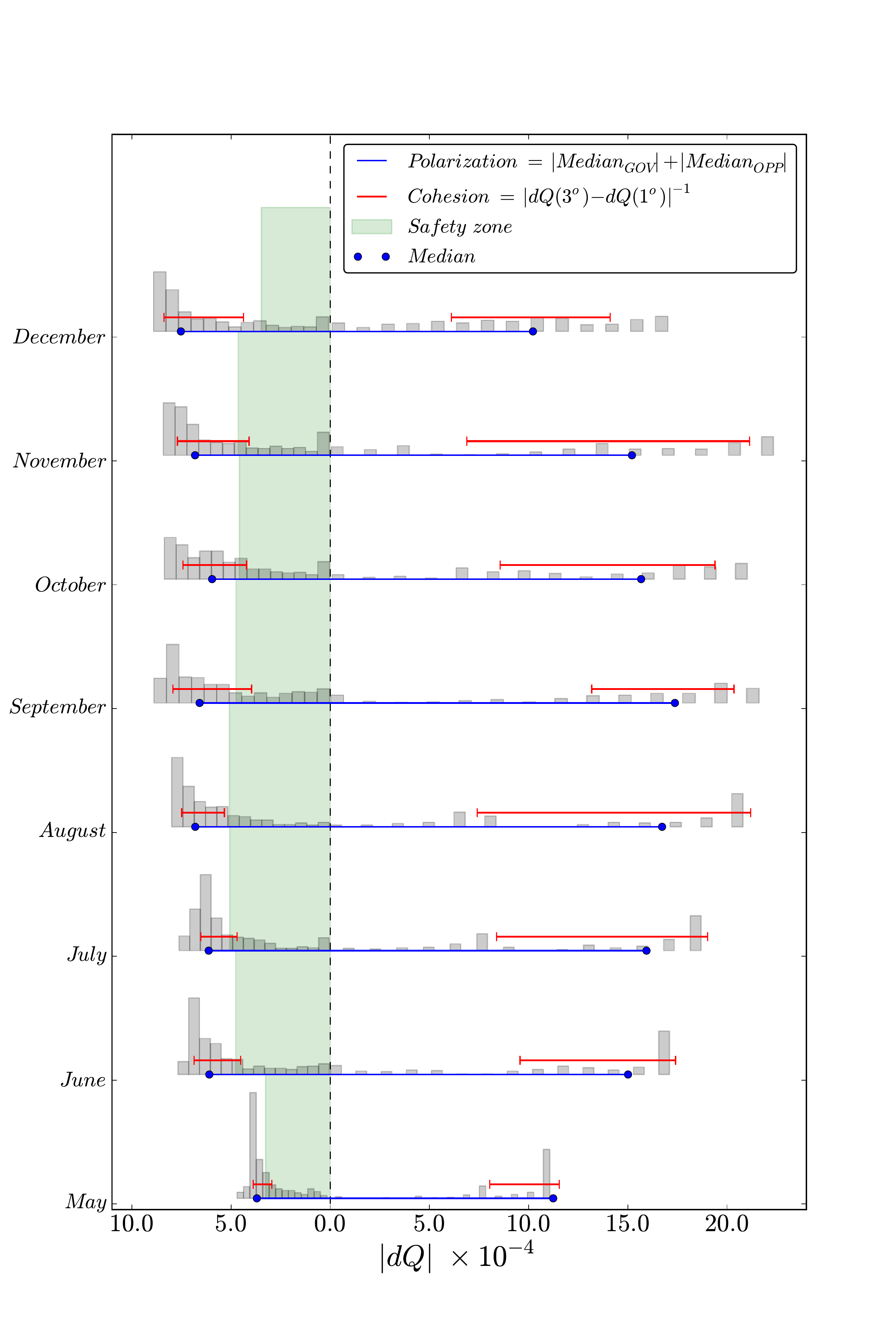}
\end{center}
\caption{
{\bf The evolution of community structure over time provides a way to track the
cohesion of the government and the overall polarization in the parliament.}  The
empyrical analog of the cohesion is represented here by the interquantile
difference of the $dQ$ distribution, where higher cohesion occurs for lower
values of the interquantile. On the other hand higher parliament polarization is
captured by the distance between the two medians. 
Finally the \textit{safety zone} that divides the last critical deputy able to
break down the majority from the vertical dashed line ($dQ=0$) is represented in
green.}
\label{Figure_3}
\end{figure}

The same analysis has been carried out over time, dividing votes per month, building up the corresponding graphs and performing the community and core analysis on each monthly network. 
In figure \ref{Figure_3} the two main communities present increasingly extreme
values of $dQ$ over time, which in turns provides evidence of increasing
polarization in the parliament.

As outlined in the Methods section we measure the polarization as the
sum of the median of the monthly $dQ$ distributions of the two communities.
In this respect, December noticeably stands out, with a reduction in the extreme values of $dQ$ for the opposition. 
This is actually driven by the fragmentation of the PDL, which witnessed its deputies loyal to the leader Silvio Berlusconi, withdraw their support of the government and start to vote with the opposition to the point of being identified as part of it.\\
This figure \ref{Figure_3} illustrates also the cohesion, or rather its flip
side:
the heterogeneity of deputies within a single community.\\
In the same figure \ref{Figure_3} we have represented the government stability
through a green \textit{safety zone} corresponding to values
of $dQ$ smaller (in absolute value) than the monthly critical value
$dQ_{critical}$. 
In the specific case of the Italian Parliament the Chamber of Deputies has $630$
representatives and the critical value will correspond to the $dQ$ relative to
the $316th$ deputy.
If the government loses this deputy will be in trouble otherwise will remain
still in the \textit{safe} zone.

\begin{figure}[!ht]
\begin{center}
\hspace*{-0.7cm}\includegraphics[width=4.0in]{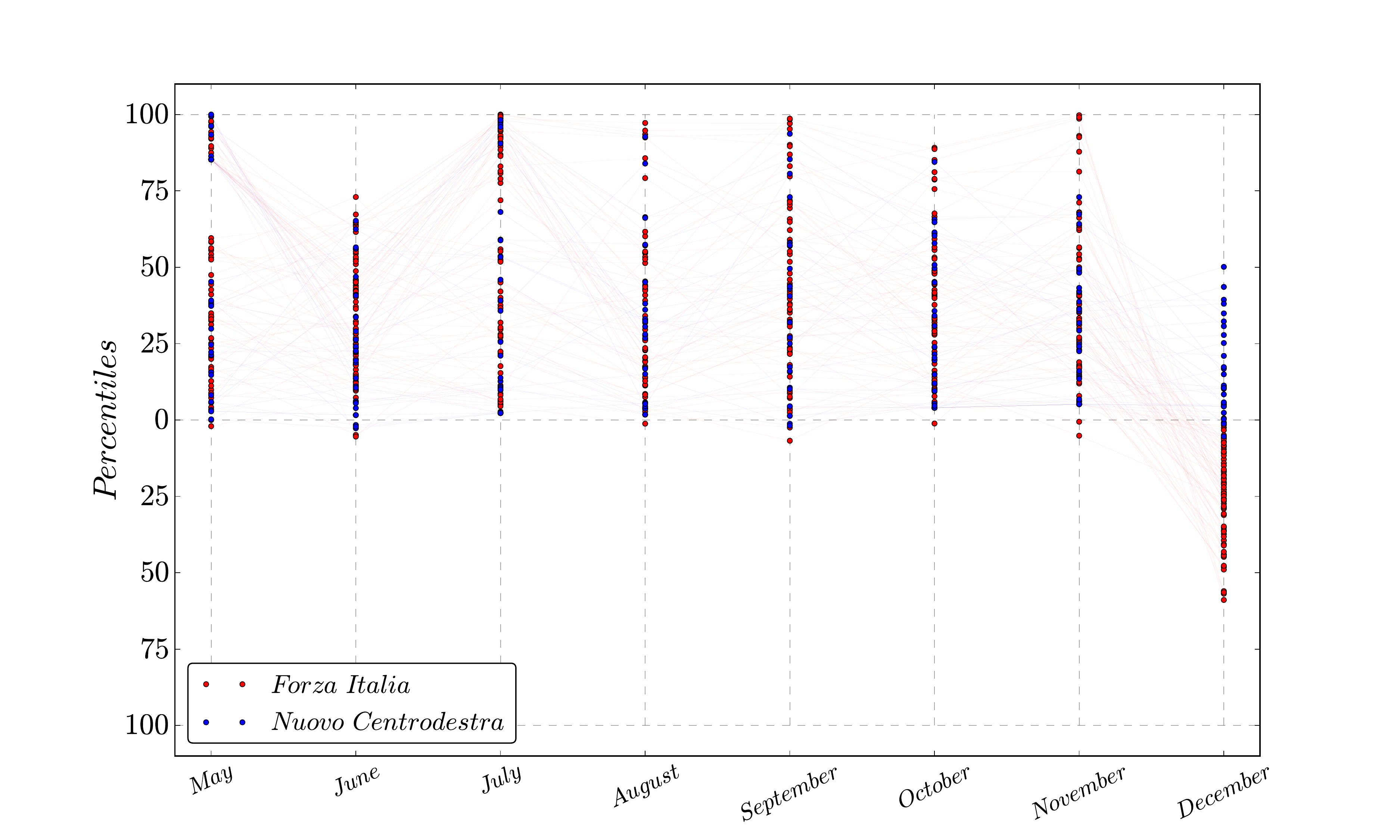}
\end{center}
\caption{
{\bf The position of single deputies within communities provides insights on what happens when a party splits up.} 
In this particular case the PDL party in mid November 2013 splits into two
different parties 'Forza Italia' and 'Nuovo Centrodestra'. Interestigly, nodes at the
core of the government coalition become core in the opposition one when the split up occurs. This is evidence of political voting being driven by coalitions' affilitions rather than the policy content of each roll call vote.}
\label{Figure_4}
\end{figure}

With fixed levels of polarization and cohesion, the wider this zone is, the
safer the government is in that month, as it would be able to retain the
withdrawal of core deputies proportionally the the stregth required to move 
the $316th$ deputy to the other side.\\
Having investigated the peculiar structure of the government coalition, we focus on a political party that may be partly responsible for the variability of the coalition�s topology over time. Indeed  the PDL, after a long debate regarding whether to support the government or not, eventually split into two different parties. After the split in mid November, deputies from the FI moved into the opposition community. 
However, surprisingly, those who left moved from the core of the government to relatively core positions in the opposition, as reported in figure \ref{Figure_4}.

This dynamic may somehow explain the peculiar drop of the polarization observed in December in figure \ref{Figure_3}, as the FI group switched voting behavior to such a degree as to be recognized as part of the opposition, simultaneously reducing the contraposition between the two communities.

 \section{Discussion}
The study of the consensus dynamics in modern parliamentary democracies is of great importance for the validation by citizens of the performance of their representatives. These dynamics are often hidden by complicated voting procedures that prevent the easy identification of these civil representatives. We need new ways to look at the details of the political activities, which go beyond the standard statistical indicators, ways that are able to reveal the dynamics of the general organization of the government, its opposition and even their internal structures, in a format that is intelligible to non-expert users. 
In this study we introduced a novel procedure to map parliamentary voting trends onto a network structure in which the nodes are the deputies and the edge weights are the strength of their relations. These weights, month by month, quantitatively measure the degree of closeness between couples of deputies as the number of votes they shared in a specific time frame. Once this network has been built up, using �Community Detection� techniques borrowed from Complex Network Science, it is possible to reconstruct the main coalitions, the government and the opposition from the bottom up; through a `Core Detection' analysis it is also possible to uncover the internal structure of these aggregations. Using the leverage of later analyses we were able to quantitatively detect the position of each party, the strength and consistency in its coalition and the level of polarization between government and opposition.\\
Furthermore, the Open Data movements around the world are pushing public administrations to provide free and open access to massive amounts of data, which can be used by citizens and companies as a starting point for the detailed analysis of public policies. In this study, we relied on a recent service introduced by the Italian parliament that allows the automated extraction of certified information about the votes of the Chamber of Deputies. Through this service we have been able to perform a thorough analysis of the dynamics of the Italian parliamentary factions over nearly a year of legislation, using the aforementioned methodology.\\
These methods open up new possibilities of bringing citizens closer to their representatives, thereby establishing the foundations for a more transparent democracy.

% Do NOT remove this, even if you are not including acknowledgments
\section*{Acknowledgments}
The authors acknowledge support from the Progetto di Interesse CNR ``CRISISLAB",
and the FET project SIMPOL 610704. All authors acknowledge enlightening discussions and support from people in LINKALAB. The funders had no role in study design, data collection and analysis, decision to publish, or preparation of the manuscript.

\begin{table}[!ht]
\caption{
\bf{Outline of the main Italian parties}}
\begin{tabular}{|c|c|c|c|}
\hline
Party & Coalition & $\%^{\ast}$ & Notes\\
\hline
Partito Democratico (PD) & Gov & $46,5 \%$& Main center-left party, historically lead by Prodi\\
\hline
Il Popolo della Libert\`{a} (PDL) & Gov/Opp & $15,2 \%$ & Main conservative party, lead by Berlusconi\\ 
\hline
Forza Italia (FI) & Opp & $10,6 \%$ & From PDL split, founded and lead by Berlusconi\\
\hline
Nuovo Centro Destra (NCD) & Gov & $4,6 \%$ & From PDL split, lead by Alfano\\
\hline
Scelta Civica (SC) & Gov & $7,3 \%$ & Lead by Monti, PM for one year after 2011 crisis\\
\hline
Movimento 5 Stelle (M5S) & Opp & $16,3 \%$ & Lead by comedian Grillo, form of \textit{direct} democracy\\
\hline
Sinistra-Ecologia-Libert\`{a} (SEL) & Opp & $5,6 \%$ & Left party, former ally of PD\\
\hline
Lega Nord (LN) & Opp & $3,2 \%$ & Autonomist party of Northern Italy, former ally of PDL\\
\hline
\end{tabular}
\begin{flushleft}
$\ast$ Shares updated to may 2014, smaller parties omitted.

\end{flushleft}
\label{tabella_carlo}
\end{table}

%\section*{References}
% The bibtex filename
\bibliography{biblio}

\begin{thebibliography}{15}
\expandafter\ifx\csname natexlab\endcsname\relax\def\natexlab#1{#1}\fi
\expandafter\ifx\csname bibnamefont\endcsname\relax
  \def\bibnamefont#1{#1}\fi
\expandafter\ifx\csname bibfnamefont\endcsname\relax
  \def\bibfnamefont#1{#1}\fi
\expandafter\ifx\csname citenamefont\endcsname\relax
  \def\citenamefont#1{#1}\fi
\expandafter\ifx\csname url\endcsname\relax
  \def\url#1{\texttt{#1}}\fi
\expandafter\ifx\csname urlprefix\endcsname\relax\def\urlprefix{URL }\fi
\providecommand{\bibinfo}[2]{#2}
\providecommand{\eprint}[2][]{\url{#2}}

\bibitem[{\citenamefont{Layman et~al.}(2006)\citenamefont{Layman, Carsey, and
  Horowitz}}]{c1}
\bibinfo{author}{\bibfnamefont{G.~C.} \bibnamefont{Layman}},
  \bibinfo{author}{\bibfnamefont{T.~M.} \bibnamefont{Carsey}},
  \bibnamefont{and} \bibinfo{author}{\bibfnamefont{J.~M.}
  \bibnamefont{Horowitz}}, \bibinfo{journal}{Annu. Rev. Polit. Sci.}
  \textbf{\bibinfo{volume}{9}}, \bibinfo{pages}{83} (\bibinfo{year}{2006}).

\bibitem[{\citenamefont{Fiorina et~al.}(2008)\citenamefont{Fiorina, Abrams, and
  Pope}}]{c2}
\bibinfo{author}{\bibfnamefont{M.~P.} \bibnamefont{Fiorina}},
  \bibinfo{author}{\bibfnamefont{S.~A.} \bibnamefont{Abrams}},
  \bibnamefont{and} \bibinfo{author}{\bibfnamefont{J.~C.} \bibnamefont{Pope}},
  \bibinfo{journal}{The Journal of Politics} \textbf{\bibinfo{volume}{70}},
  \bibinfo{pages}{556} (\bibinfo{year}{2008}).

\bibitem[{\citenamefont{Caldarelli}(2007)}]{Caldarelli2007Scale-FreeNetworksComplex}
\bibinfo{author}{\bibfnamefont{G.}~\bibnamefont{Caldarelli}},
  \emph{\bibinfo{title}{{Scale-Free Networks: Complex Webs in Nature and
  Technology}}} (\bibinfo{publisher}{Oxford University Press, USA},
  \bibinfo{year}{2007}).

\bibitem[{\citenamefont{Newman}(2003)}]{newman2003structure}
\bibinfo{author}{\bibfnamefont{M.~E.} \bibnamefont{Newman}},
  \bibinfo{journal}{SIAM review} \textbf{\bibinfo{volume}{45}},
  \bibinfo{pages}{167} (\bibinfo{year}{2003}).

\bibitem[{\citenamefont{Porter et~al.}(2005)\citenamefont{Porter, Mucha,
  Newman, and Warmbrand}}]{c3}
\bibinfo{author}{\bibfnamefont{M.~A.} \bibnamefont{Porter}},
  \bibinfo{author}{\bibfnamefont{P.~J.} \bibnamefont{Mucha}},
  \bibinfo{author}{\bibfnamefont{M.~E.} \bibnamefont{Newman}},
  \bibnamefont{and} \bibinfo{author}{\bibfnamefont{C.~M.}
  \bibnamefont{Warmbrand}}, \bibinfo{journal}{Proceedings of the National
  Academy of Sciences of the United States of America}
  \textbf{\bibinfo{volume}{102}}, \bibinfo{pages}{7057} (\bibinfo{year}{2005}).

\bibitem[{\citenamefont{Fowler}(2006)}]{c4}
\bibinfo{author}{\bibfnamefont{J.~H.} \bibnamefont{Fowler}},
  \bibinfo{journal}{Political Analysis} \textbf{\bibinfo{volume}{14}},
  \bibinfo{pages}{456} (\bibinfo{year}{2006}).

\bibitem[{\citenamefont{Tam~Cho and Fowler}(2010)}]{c5}
\bibinfo{author}{\bibfnamefont{W.~K.} \bibnamefont{Tam~Cho}} \bibnamefont{and}
  \bibinfo{author}{\bibfnamefont{J.~H.} \bibnamefont{Fowler}},
  \bibinfo{journal}{The Journal of Politics} \textbf{\bibinfo{volume}{72}},
  \bibinfo{pages}{124} (\bibinfo{year}{2010}).

\bibitem[{\citenamefont{Waugh et~al.}(2009{\natexlab{a}})\citenamefont{Waugh,
  Pei, Fowler, Mucha, and Porter}}]{c6}
\bibinfo{author}{\bibfnamefont{A.~S.} \bibnamefont{Waugh}},
  \bibinfo{author}{\bibfnamefont{L.}~\bibnamefont{Pei}},
  \bibinfo{author}{\bibfnamefont{J.~H.} \bibnamefont{Fowler}},
  \bibinfo{author}{\bibfnamefont{P.~J.} \bibnamefont{Mucha}}, \bibnamefont{and}
  \bibinfo{author}{\bibfnamefont{M.~A.} \bibnamefont{Porter}},
  \bibinfo{journal}{arXiv preprint arXiv:0907.3509}
  (\bibinfo{year}{2009}{\natexlab{a}}).

\bibitem[{\citenamefont{Zhang et~al.}(2008)\citenamefont{Zhang, Friend, Traud,
  Porter, Fowler, and Mucha}}]{c7}
\bibinfo{author}{\bibfnamefont{Y.}~\bibnamefont{Zhang}},
  \bibinfo{author}{\bibfnamefont{A.}~\bibnamefont{Friend}},
  \bibinfo{author}{\bibfnamefont{A.~L.} \bibnamefont{Traud}},
  \bibinfo{author}{\bibfnamefont{M.~A.} \bibnamefont{Porter}},
  \bibinfo{author}{\bibfnamefont{J.~H.} \bibnamefont{Fowler}},
  \bibnamefont{and} \bibinfo{author}{\bibfnamefont{P.~J.} \bibnamefont{Mucha}},
  \bibinfo{journal}{Physica A: Statistical Mechanics and its Applications}
  \textbf{\bibinfo{volume}{387}}, \bibinfo{pages}{1705} (\bibinfo{year}{2008}).

\bibitem[{\citenamefont{Waugh et~al.}(2009{\natexlab{b}})\citenamefont{Waugh,
  Pei, Fowler, Mucha, and Porter}}]{c8}
\bibinfo{author}{\bibfnamefont{A.~S.} \bibnamefont{Waugh}},
  \bibinfo{author}{\bibfnamefont{L.}~\bibnamefont{Pei}},
  \bibinfo{author}{\bibfnamefont{J.~H.} \bibnamefont{Fowler}},
  \bibinfo{author}{\bibfnamefont{P.~J.} \bibnamefont{Mucha}}, \bibnamefont{and}
  \bibinfo{author}{\bibfnamefont{M.~A.} \bibnamefont{Porter}},
  \bibinfo{journal}{arXiv preprint arXiv:0907.3509}
  (\bibinfo{year}{2009}{\natexlab{b}}).

\bibitem[{\citenamefont{Newman and Girvan}(2004)}]{c11}
\bibinfo{author}{\bibfnamefont{M.~E.} \bibnamefont{Newman}} \bibnamefont{and}
  \bibinfo{author}{\bibfnamefont{M.}~\bibnamefont{Girvan}},
  \bibinfo{journal}{Physical review E} \textbf{\bibinfo{volume}{69}},
  \bibinfo{pages}{026113} (\bibinfo{year}{2004}).

\bibitem[{\citenamefont{Fortunato}(2010)}]{c10}
\bibinfo{author}{\bibfnamefont{S.}~\bibnamefont{Fortunato}},
  \bibinfo{journal}{Physics Reports} \textbf{\bibinfo{volume}{486}},
  \bibinfo{pages}{75} (\bibinfo{year}{2010}).

\bibitem[{\citenamefont{Newman}(2004)}]{newman2004analysis}
\bibinfo{author}{\bibfnamefont{M.~E.} \bibnamefont{Newman}},
  \bibinfo{journal}{Physical Review E} \textbf{\bibinfo{volume}{70}},
  \bibinfo{pages}{056131} (\bibinfo{year}{2004}).

\bibitem[{\citenamefont{De~Leo et~al.}(2013)\citenamefont{De~Leo, Santoboni,
  Cerina, Mureddu, Secchi, and Chessa}}]{c12}
\bibinfo{author}{\bibfnamefont{V.}~\bibnamefont{De~Leo}},
  \bibinfo{author}{\bibfnamefont{G.}~\bibnamefont{Santoboni}},
  \bibinfo{author}{\bibfnamefont{F.}~\bibnamefont{Cerina}},
  \bibinfo{author}{\bibfnamefont{M.}~\bibnamefont{Mureddu}},
  \bibinfo{author}{\bibfnamefont{L.}~\bibnamefont{Secchi}}, \bibnamefont{and}
  \bibinfo{author}{\bibfnamefont{A.}~\bibnamefont{Chessa}},
  \bibinfo{journal}{Physical Review E} \textbf{\bibinfo{volume}{88}},
  \bibinfo{pages}{042810} (\bibinfo{year}{2013}).

\bibitem[{\citenamefont{Moody and Mucha}(2013)}]{c9}
\bibinfo{author}{\bibfnamefont{J.}~\bibnamefont{Moody}} \bibnamefont{and}
  \bibinfo{author}{\bibfnamefont{P.~J.} \bibnamefont{Mucha}},
  \bibinfo{journal}{Network Science} \textbf{\bibinfo{volume}{1}},
  \bibinfo{pages}{119} (\bibinfo{year}{2013}).

\end{thebibliography}

%\section*{Tables}

\end{document}